\documentclass{webofc}
\begin{document}
%
%\title{Recent Developments in Hydrodynamics}
\title{Conservation Laws and Hydrodynamics}

\author{\firstname{Sangyong}
\lastname{Jeon}\inst{1,3}\fnsep\thanks{\email{jeon@physics.mcgil.ca}} 
}

\institute{
Department of Physics, McGill University, 3600 rue University, Montreal, QC, Canada
          }

\abstract{%
Recent years have seen much development in analyzing the structure of relativistic
hydrodynamics. In this proceeding, some of the developments are highlighted including
issues related to pseudo-gauge transformations and
spin hydrodynamics.
}
\maketitle
\section{Introduction} \label{intro}

Hydrodynamics has a long history in describing heavy ion collisions 
starting with Landau's seminal paper \cite{Landau1953} and Bjorken's
\cite{Bjorken:1982qr}. In the era of RHIC and the LHC, hydrodynamics
has become an indispensable tool to describe the bulk dynamics of 
relativistic heavy ion collisions.

In this 2022 Strangeness in Quark Matter conference, many of the recent developments in
hydrodynamics were
reported. On spin hydrodynamics and $\Lambda$ polarization, Palermo \cite{Palermo:2022lvh},
Liao \cite{Guo:2021udq}
and Buzzegoli \cite{Buzzegoli:2022kyj} presented their work. 
On the subject of nonolinear causality conditions, Plumberg \cite{Plumberg:2021bme} presented
his and collaborators' work. On the development of hydro initial states, 
Shen \cite{Zhao:2022ayk} presented recent developments for ultra-peripheral collisions and 
Kanakubo \cite{Kanakubo:2022acq} presented her recent work on the core-corona model. 
On applying hydrodynamics to systems near the QCD critical point, 
Pradeep \cite{Pradeep:2022mkf}, Wu \cite{Wu:2021fjf}, and Sogabe \cite{Sogabe:2021svv} presented 
their work in this conference. 

As this is a brief introduction to the recent developments, inevitably
only a limited number of topics can be discussed. Below, I will mainly
focus on the pseudo-gauge transformation and its relevance in developing spin hydrodynamics.
I should emphasize that 
this is a personal choice dictated by my familiarity with the subjects \cite{Shi:2020htn}. 

\section{Is any current density observable?}
\label{sec-1}

Hydrodynamics starts with conservation laws
\begin{equation}
\partial_\mu J^{\mu,q} = 0
\end{equation}
where $q$ is the index for different conserved quantities such as energy, momentum, electric charge
and baryon number.
Mathematically, a conserved current density is {\em not} uniquely defined.
One can always add a ``pseudo-gauge'' term
\begin{equation}
{J'}^{\mu,q} = J^{\mu,q} + \partial_\lambda B^{\lambda\mu,q}
\label{eq:pseudo_gauge}
\end{equation}
where $B^{\lambda\mu,q} = -B^{\mu\lambda,q}$ is the pseudo-gauge potential.
Adding $\partial_\lambda B^{\lambda\mu,q}$ neither changes the conservation law
$\partial_\mu {J'}^{\mu,q} = 0$, nor the total charge
provided that $B^{i0,q}$ vanishes at the boundary of the volume.
But it does change the local current density since $J^{\mu,q}(x) \ne {J'}^{\mu,q}(x)$.

Does this mean that a current density is not observable? 
That depends on whether there exist other
constraints that may select a particular form to be physical.
For instance, in the classical particle dynamics, a point particle carries well-defined 
energy, momentum and other conserved charges. Hence, the classical current density
\begin{equation}
J_{\rm cl}^{\mu,q}(x) = \sum_{s=\rm ptcles} {q_s\over p_s^0} p^\mu_s \delta^{(3)}({\bf x} - {\bf x}_s(t))
\label{eq:classical_J}
\end{equation}
is physical and observable as long as a particle is observable.
Adding an arbitrary $\partial_\lambda B^{\lambda\mu,q}$ to $J^{\mu,q}_{\rm cl}$
is still a possibility,
but the fact that a point particle carries definite amounts of
energy, momentum and other charges clearly selects
Eq.(\ref{eq:classical_J}) to be the physical current density.

In reality, 
due to the uncertainty principle, 
particles cannot be localized as in Eq.(\ref{eq:classical_J}).
Particle positions are inevitably smeared
so that
\begin{equation}
J_C^{\mu,q}(x) = \sum_{s=\rm ptcles} {q_s\over p_s^0} p^\mu_s g({\bf x} - {\bf x}_s(t))
\end{equation}
with the normalization
condition $\int d^3x\, g({\bf x} - {\bf x}_s(t)) = 1$ that preserves the total charge.
This expression still satisfies $\partial_\mu J_C^{\mu,q} = 0$ as long as $d{\bf x}_s/dt = {\bf p}_s/p^0_s$,
but since there can be many choices for the smearing function,
this does not define a unique value
of the current density at $x$.
For any change $\delta{J}^{\mu,q}_C$ introduced by a change of the smearing function,
one can always find $B^{\lambda\mu,q}$ that satisfies
\begin{equation}
\delta{J}^{\mu,q}_C = \partial_\lambda B^{\lambda\mu,q}
\label{eq:deltaJ}
\end{equation}
since it has the same form as the Maxwell equation. Hence, without any further constraint on the form of the
smearing function, the continuous current density $J_C^{\mu,q}$ can be ambiguous.

I have been so far 
careful to always refer to $J^{\mu,q}$ as a current density. This is to distinguish $J^{\mu,q}$ from
the current. An electric current, for instance, is defined as
\begin{equation}
I_S = \int_S d{\bf S}\cdot {\bf J}^e
\end{equation}
where ${\bf S}$ is the surface through which the electric current is measured and ${\bf J}^e$ is the spatial part
of the electric current density.
The current itself is a scalar quantity, not a vector quantity.
One can easily show that 
as long as ${\bf B}$ vanishes at the boundary of the surface, $I_S$ remains unchanged.

Does this mean that no current density is actually observable 
due to the pseudo-gauge freedom?
In order to answer this question,
consider the electric current density $J_e^\mu$.
As discussed above, the conservation law is not affected when
a pseudo-gauge potential with the property
$B^{\nu \mu}_e = -B^{\mu\nu}_e$ is added to $J_e^\mu$.
Using the new current density,
The Maxwell equation becomes
\begin{equation}
\partial_\nu {F'}^{\nu\mu} = {J'_e}^{\mu} = J_e^{\mu}  + \partial_\nu B_e^{\nu\mu}
\end{equation}
The electric and magnetic field generated by ${J'_e}^\mu$ is then
\begin{equation}
{F'}^{\nu\mu} = F^{\nu\mu} + B_e^{\nu\mu}
\end{equation}
where $F^{\nu\mu}$ is generated by $J_e^{\mu}$.
But this doesn't make much sense. The electric field and the magnetic field are
measurable observables. They shouldn't be ambiguous.
Hence, there must be a {\em physical} electric current density 
that generates the {\em measured} electric and magnetic fields.

For the energy-momentum tensor, a similar situation arises with the Einstein's equation where a symmetric
Belinfante energy-momentum tensor (see below)
sources the gravitational field. Since gravitational field is observable and measurable, 
the Belinfante tensor 
must be the physical energy-momentum tensor.

There are two lessons I would take from the discussions above.
One is that as long as a current density generates an observable
and measurable field (such as the electric and magnetic fields),
one should be able to identify the unique physical and observable current density.
The second lesson is that even though unambiguously
identifying {\em classical} current density
using the above arguement may be  possible, 
identifying quantum current density could be still ambiguous (c.f.~Eq.(\ref{eq:deltaJ})).
However, this ambiguity should represent only the
$O(\hbar)$ correction to the classical current density.

\section{Stress-energy tensors with spin}

Hydrodynamics primarily deals with the energy-momentum tensor.
Consider a field-theory Lagrangian $L(\varphi^A, \partial_\mu\varphi^A)$ where $A$ is the field index.
Using Noether's theorem, one can show that the conserved current densities for the 
translational symmetry in space-time is given by 
\begin{equation}
\Theta^{\mu\nu} = \Pi_A^\mu \partial^\nu\varphi^A - g^{\mu\nu}L(\varphi_A, \partial_\mu\varphi^A)
\label{eq:canonical_Tmunu}
\end{equation}
where $\Pi_A^\mu = \partial L/\partial(\partial_\mu\varphi_A)$.
This tensor satisfies
\begin{align}
\partial_\mu\Theta^{\mu\nu} = \partial^\nu\varphi_A
\left(
\partial_\mu \Pi^\mu_A - {\partial L\over \partial \varphi_A}
\right)
\end{align}
Hence if $\varphi_A$ satisfies the Euler-Lagrange equation
$\partial_\mu \Pi^\mu_A = \partial L/ \partial \varphi_A$, 
then $\Theta^{\mu\nu}$ is a conserved energy-momentum current density.
The first index $\mu$ is the space-time index for the current density components
and the second index $\nu$ indicates what is being conserved.
Namely, $\nu = 0$ is the label for the energy current
and $\nu = 1, 2, 3$ are for the momentum currents playing the role of
the index $q$ in the previous section.
This canonical energy-momentum tensor 
is not symmetric unless it only contains scalar (spin $S = 0$) particles, and
it is not gauge invariant when applied to gauge theories.
As such, it cannot yet be directly interpreted
as the measurable energy-momentum current density.

To see the consequences of having a non-symmetric $\Theta^{\mu\nu}$,
define the orbital angular momentum current density
\begin{equation}
{\cal L}_O^{\lambda\mu\nu} = x^\mu \Theta^{\lambda\nu} - x^\nu \Theta^{\lambda\mu}
\end{equation}
This tensor is in general not conserved
$\partial_\lambda {\cal L}_O^{\lambda\mu\nu} = \Theta^{\mu\nu} - \Theta^{\nu\mu}$
unless all the fields have $S = 0$.
If fields have non-zero spins, then the total angular momentum 
current density
must include the spin angular momentum part as well as the orbital angular momentum part
\begin{equation}
{\cal L}_{\rm tot}^{\lambda\mu\nu} = 
 x^\mu \Theta^{\lambda\nu} - x^\nu \Theta^{\lambda\mu} + {\cal S}^{\lambda\mu\nu}
\end{equation}
where ${\cal S}^{\lambda\mu\nu}$ is the spin current density.
The statement of the total angular momentum conservation becomes
\begin{equation}
\partial_\lambda {\cal S}^{\lambda\mu\nu} + 2\Theta_{(a)}^{\mu\nu} = 0
\label{eq:ang_cons}
\end{equation}
In other words, the anti-symmetric part of the energy-momentum tensor,
$\Theta_{(a)}^{\mu\nu} = (1/2)(\Theta^{\mu\nu} - \Theta^{\nu\mu})$,
encodes the spin angular momentum part of the
energy and momentum. That part is then necessarily of $O(\hbar)$ 
since the spin current density is $O(\hbar)$.

%There is, however, a further complication. We can add a pseudo-gauge potential term $\partial_\lambda
%B^{\lambda\mu,\nu}$ to $\Theta^{\mu\nu}$ without violating energy-momentum conservation.
%We can also add to the spin current density another pseudo-gauge potential term
%\begin{align}
%{\cal S'}^{\lambda\mu\nu} = {\cal S}^{\lambda\mu\nu} + \partial_\rho Z^{\rho\lambda,\mu\nu}
%\end{align}
%such that $Z^{\rho\lambda,\mu\nu} = -Z^{\lambda\rho,\mu\nu}$ 
%and $Z^{\rho\lambda,\mu\nu} = -Z^{\rho\lambda,\nu\mu}$ 
%without affecting Eq.(\ref{eq:ang_cons}).

To find the pseudo-gauge transformation that results in the physical
energy-momentum tensor,
let
\begin{align}
T^{\mu\nu} = \Theta^{\mu\nu} + \partial_\lambda B^{\lambda\mu,\nu}
\label{eq:Tmunu}
\end{align}
where $T^{\mu\nu}$ is the symmetric tensor that we are after.
Using $T^{\mu\nu} - T^{\nu\mu} = 0$, we get
\begin{align}
0 & = 
\Theta^{\mu\nu} - \Theta^{\nu\mu} + \partial_\lambda(B^{\lambda\mu,\nu} - B^{\lambda\nu,\mu})
\label{eq:pseudo_gauge_1}
\end{align}
Using the conservation of total angular momentum, Eq.(\ref{eq:ang_cons}), one can identify
\begin{align}
{\cal S}^{\lambda\mu\nu} = 
B^{\lambda\mu,\nu} - B^{\lambda\nu,\mu}
\end{align}
Equivalently,
\begin{align}
B^{\lambda\mu,\nu}
=
{1\over 2}\left(
{\cal S}^{\lambda\mu\nu} - {\cal S}^{\mu\lambda\nu} - {\cal S}^{\nu\lambda\mu}
\right)
\label{eq:B3}
\end{align}
that yields
\begin{align}
T^{\mu\nu}
& = 
\Theta^{\mu\nu}_{(s)}
- {1\over 2}\partial_\lambda\left( {\cal S}^{\mu\lambda\nu} + {\cal S}^{\nu\lambda\mu} \right)
\label{eq:Tmunu_with_spin}
\end{align}
after explicit symmetrization.
This form of the energy-momentum tensor is usually referred to 
as the Belinfante energy-momentum tensor
and it can be shown that this is identical to the gravitational source term 
in the Einstein equation
obtained by varying the space-time metric.
As such, classically, this is the observable energy-momentum tensor.
One should note that this form of energy-momentum tensor
now incorporates the conservation of total angular momentum 
since $\partial_\mu T^{\mu\nu} = 0$ is satiafied
only after using both
the equations of motion and the total angular momentum
conservation, Eq.(\ref{eq:ang_cons}). 
It also means that the energy density obtained from $T^{\mu\nu}$ 
contains the energy due to the spins \cite{Fukushima:2020ucl}.

The the total angular momentum current density is now just given by
\begin{align}
{\cal L}_{\rm tot}^{\lambda\mu\nu} = 
 x^\mu T^{\lambda\nu} - x^\nu T^{\lambda\mu} 
\end{align}
which could be 
further divided this into the orbital angular momentum term 
and the spin term although this division is not so 
clear-cut.\footnote{For instance, for a spin-1 gauge field, only the sum is gauge-invariant.}
This also means that there is no separate spin current density which presents a bit of a
problem when formulating spin hydrodynamics.

\section{Spin hydrodynamics}

STAR 
has measured $\Lambda$ polarizations that showed a surprising azimuthal pattern
\cite{STAR:2017ckg} that may
indicate a strong spin-orbit coupling between the polarization of a hadron and the underlying fluid
motion. In order to figure out where such a strong correlation can originate
from, it is clear that we need
a formulation of hydrodynamics that includes the spin degrees of freedom. This is, however, not
straightforward:
If one insists that the energy-momentum tensor must be observable, 
then it must be the Belinfante tensor.
But that means there is no separate spin current density.
On the other hand,
one can naturally get
the spin current density via the anti-symmetric part of the canonical energy-momentum tensor.
However, the canonical energy-momentum tensor is not physical.

There are currently two broad approaches to this problem. One is to approach the spin
hydrodynamics from the semi-classical kinetic theory point of view 
\cite{Palermo:2022lvh,Buzzegoli:2022kyj,Weickgenannt:2022zxs,Shi:2020htn}.
Another approach is to start with the
general expression of the non-symmetric energy-momentum tensor and the spin current density
tensor and develope a gradient expansion \cite{Daher:2022xon,Fukushima:2020ucl}.
One can then use the extended thermodynamic identities and the production of entropy to
obtain the first order and the second order form of spin hydrodynamics.
The relationships between various
formulations of the spin hydrodynamics, however, is not yet fully settled
and full spin hydrodynamics simulations are yet to be performed.

In all these approaches, a common task is to decompose the non-symmetric tensor into the
physical symmetric part and the $O(\hbar)$ spin dependent part.
This decomposition is necessary because quantities such as the energy density and the flow
velocity must be defined to carry out any hydrodynamic calculations.
In some sense, this largely deals with the pseudo-gauge problem. Namely, once
we have identified
the classical energy-momentum tensor (corresponding to the Belinfonte tensor), then 
only $O(\hbar)$ pseudo-gauge potentials are allowed in order not to spoil the classial part
of the energy-momentum tensor.

\section{Outlook}

In this short review, I have tried to highlight some of the issues in formulating physical
hydrodynamics that were raised in recent literatures.
The two related topics I have chosen to
highlight here, namely the issue of pseudo-gauge transformations and spin hydrodynamics, have 
been studied by several groups motivated by the $\Lambda$ polarizations measured by
the STAR collaboration at RHIC. 
Through such efforts, it has become clearer what role pseudo-gauge transformations 
can and cannot play in hydrodynamics in general and in spin hydrodynamics in particular.

Theoretically, the STAR $\Lambda$ polarization measurement presents an interesting challenge.
Although some resolutions of this puzzle\footnote{The measured azimuthal dependence of the
$\Lambda$ polarization has the opposite sign compared to the vorticity tensor.} have been
suggested \cite{Palermo:2022lvh}, more studies, especially realistic spin hydrodynamics
simulations, are needed to firmly settle it.

\section*{Acknowledgement}
S.J. is supported in part by the Natural Sciences and
Engineering Research Council of Canada.

%
% BibTeX or Biber users please use (the style is already called in the class, ensure that the "woc.bst" style is in your local directory)
% \bibliography{name or your bibliography database}

\end{document}